\documentclass{amsart}
\newenvironment{nouppercase}{%
  \renewcommand{\uppercasenonmath}[1]{}}{}
\usepackage{enumerate,mathrsfs}
\usepackage{amsmath}
\usepackage{braket}
\usepackage{color}
\usepackage{graphicx}
\usepackage{tikz}
\usetikzlibrary{calc, shapes, graphs, positioning,backgrounds,arrows,automata,backgrounds,fit,decorations.pathreplacing,hobby,celtic,knots,decorations.pathreplacing,shapes.geometric,intersections
}

\DeclareMathOperator{\rank}{rank}
\DeclareMathOperator{\spn}{span}

\theoremstyle{definition}
\newtheorem{defn}{Definition}
\newtheorem{example}{Example}
\newtheorem{theorem}{Theorem}
\newtheorem{lemma}{Lemma}

\newcommand\bigsubset[1][1.19]{
   \mathrel{\vcenter{\hbox{\scalebox{#1}{$\subset$}}}}}

\title{\text{Subfactors from graphs induced by association schemes}}
\author{Radhakrishnan Balu$^{\dag}$}
\address{$^{\dag}\text{Army Research Office,
RTP, NC, 21005-5069, USA}$.}
\email{radhakrishnan.balu.civ@army.mil}

\begin{document}
\begin{nouppercase}
\maketitle
\end{nouppercase}
\begin{abstract}
We characterize anyonic systems algebraically by identifying the mathematical structures that support duality and fusion, Reidemeister moves, that are invariants of knots, braids, and modular data. The characterization is based on the connection between fusion algebras relevant in conformal field theories and character algebras related to association schemes. To make this abstract connection concrete, we provide the example of Hamming association schemes and relate them to representations of quantum groups $SU_q(2)$ that are closely connected to $SU(2)_k$ algebras whose fusion rules describe well known anyons. Our primary object of interest is the interacting Fock space which is deeply connected to an association scheme and the corresponding Bose-Mesner algebra, a combinatorial gadget with built-in duality and fusion rules, that leads to matrices (invariant under Reidemeister II and III moves in knots) which aid construction of subfactors with projections that braid. This way we set up a subfactor, a 3D topological quantum field theory, and a 2D rational conformal field theory and relate them to anyon systems described by fusion algebras. We discuss in detail a large family of graphs of self-dual association schemes that can be treated with this algebraic framework. 
\end{abstract}
\section {Introduction}
Fault-tolerant quantum computation models are based on quasiparticles called anyons that obey exchange statistics between that of fermions and bosons. They arise as excitations of two dimensional systems such as the fractional quantum Hall effect (FQHE) and discrete gauge theories. Theoretical descriptions of anyons are based on the correspondence between 2D rational conformal theories (RCFT) and 3D topological field theories (TFT). According to this correspondence, the ground state Laughlin wavefunctions of FQHE are described by conformal blocks of RCFT 
that are equivalent to the 3D states of the space-time manifold described by TFT \cite {MooreRead1991}. There is a large set of models that display anyon based CFT-TFT correspondence that shed further insights into highly correlated systems such as the FQHE \cite {Gromov2017}. 

The mathematical description of topological quantum computation is carried out at several levels \cite {Rowell2012}. The unitary modular tensor categories (UMC) form the bulk of the literature \cite {Bruillard2012}, \cite {Bruillard2016} as they classify realizable anyon systems for topological quantum computation. As this form of quantum computation is based on topological quantum field theory \cite {Pachos2012}, like the SU(2) Chern-Simons theory, the cobordism hypothesis \cite {Freed2013} provides another level of description of the underlying processes especially on manifolds with boundaries, corners, and anomalies. From the algebraic point of view, von Neumann algebras of type $II_1$ play an important role in describing topological quantum computation. Specifically, subfactors relate to higher order Braid groups and knot (Wilson loops of anyonic worldlines) invariants that describe the interacting anyons to encode unitaries via their representations in Temperley-Lieb (TL) algebras. In this manuscript, we describe the ways the generators of $TL_n$ naturally arise from association scheme induced type-II matrices. Subfactors are type $II_1$ factors, with trivial centers, von Neumann algebras with a trace that maps projections to the interval $[0, 1]$. In a sense they describe continuous geometries, and one simple way to construct type $II_1$ factors is using Clifford algebras. We can start with complex numbers and embed them diagonally onto $2 \times 2$ matrices and further embed theses matrices diagonally into $4 \times 4$ matrices and so on. The resulting sequence of Clifford algebras, with normalized traces by a factor of $\frac {1} {2^n}$ at each stage, leads in the limit to a type $II_1$ factor called hyperfinite. When we have a type $II_1$ factor $A$ as a sub *-algebra of another type $II_1$ factor $B$, we call $A \subset B$ a subfactor. In the case of subfactors, the projections between the stages of the sub *-algebras obey the braiding rules. 

For an equivalence between subfactors and UMC in describing anyon systems, we refer the readers to the work of Wang et al \cite {Wang2008} where they derive a TL from an UMC for a {'}golden chain{'} of interacting Fibonacci anyons. The {'}golden chain{'} is constructed by truncating irreducible representations of the compact group SU(2). The {'}quantized{'} SU(2) at level $k$ has elements that satisfy fusion rules where Ising anyons have level $k=2$ and Fibonacci anyons have $k=3$. Wilson loops are defined in terms of the trace of path integrals with action terms contain the elements of SU(2) \cite {Pachos2012}. Thus, the expectation of Wilson loops (knots and links) are the same as the expectation of the anyon evolution. To have the expectation of the Wilson loop invariant with respect to continuous deformations, it has to be compatible with Reidemeister moves of crossings. With the equivalence of knots and Wilson loops, we can carry out the analysis of the combinatorics in terms of algebra. One way to understand this is to look at the spins, with the symmetry described by $SU(2)$, that have irreducible representation (IRR) whose members when tensored (composite particles) are no longer irreducible. The tensor product can then be decomposed into IRR with Clebsch-Gordon coefficients and this {'}angular momentum{'} has the fusion rules similar to interacting anyons. The algebra of truncated $SU(2)_k$ with fusion rules is Temperley-Lieb which leads to a subfactor, \cite {Kaul1994} and more general topological systems with colored braids and multicolored links are conceivable by this connection. This follows when we look at the interaction Hamiltonian between anyons as the generalization of Heisenberg spin chains \cite {Wang2008}.
\begin {align*}
\mathcal{H}^{SU(2)}_{Heisenberg} &= J\sum_{<ij>} \vec{S_i}\cdot\vec{S_j} = \frac{J}{2}\left(\sum_{<ij>}\Pi_{ij}^0 - \frac{3}{2} \right). \\
\mathcal{H}^{SU(2)_k}_{Heisenberg} &= J\sum_{<ij>} \Pi_{ij}^1.
\end {align*}
The projection $\Pi_{ij}^0$ which encodes the antiferromagnetic ground state in Heisenberg spin chains has a counterpart in the truncated version which is the projection $\Pi_{ij}^1$ that satisfies the Temperley-Lieb relations. The derivation of this projector is complex even for the simplest case of Fibonacci anyons because of the combinations of fusion paths to consider. This process simplifies when we start with $SU(2)_k$ Wess-Zumino conformal field theory and build a representation of braid group followed by a construction of a subfactor and the Temperley algebra. We will describe a similar construction from fusion rules of association schemes to braid group representation and build the same subfactors TL algebras. For the Hamming association scheme, we will embed its Bose-Mesner algebra into a group whose representation is related to $SU(2)_k$.
We will set up type II matrices that encode crossings of knots, anyon interactions, from a Bose-Mesner algebra of a self-dual  association scheme and describe the induced subfactor and the associated projectors with examples. 

In formal terms, let $U^\chi$ be a unitary irreducible representation of a compact (second countable) group G, like $SU(2)$, with character $\chi \in \Gamma(G)$ (dual of G) with dimension $d(\chi)$. If $\chi_1, \chi_2 \in \Gamma(G)$ then the tensor product $U^{\chi_1} \otimes U^{\chi_2}$ decomposes into a sum of IRR. Let us denote by $m(\chi_1, \chi_2; \chi)$ the multiplicity with which type $U^\chi$ appears in such a decomposition of  $U^{\chi_1} \otimes U^{\chi_2}$. By letting $$ p^{\chi}_{\chi_1, \chi_2} = \frac {m(\chi_1, \chi_2; \chi)d(\chi_2)}{d(\chi)d(\chi_1)} $$ we have the Clebsch-Gordon formula
\begin {align*}
p^{\chi}_{\chi_i, \chi_j} &= \frac{i - 1}{2i}, \text {if  } j = i - 1, \\ 
p^{\chi}_{\chi_i, \chi_j} &= \frac{i + 1}{2i}, \text {if  } j = i + 1, \\
&= 0, \text {  otherwise.}
\end {align*}

In our work, we generate the fusion and braiding from association schemes. From the fusion rules of the association scheme, we can derive the $S-$matrix, with eigen vectors of the scheme forming the columns, and the Verlinde formula for the anyons. There is a one-to-one correspondence between character algebras related to association schemes and fusion algebras of conformal field theories such as the $SU(2)_k$ based ones\cite {Bannai1993} as axiomatic frameworks.  Gannon elaborated this correspondence and discussed several fusion algebras including association schemes in the language of conformal field theory \cite {Gann2005}. The key idea is the modular data or invariance present in these systems that is described in terms of the $S-$matrix as $(ST)^2 = S^2, T \text { a diagonal matrix}$. Gannon further considered graphs that are representations of fusion rules in terms of their adjacency matrices as the Cartan matrices of $SU(2)_k$ whose fusion rules correspond to Dynkin graphs. He observed that the association schemes have a large classification compared to modular data, and the benefits to probe further the connection to association schemes. We will provide another association by embedding the Bose-Mesner algebra into a quantum group. Nomura established the modular data property for association schemes of distance-regular graphs that are not bipartite \cite {Nomura2002}. We focus on such graphs in this paper as we explained in our earlier work on distance-regular graphs \cite {Radbalu2020, Radbalu2021}, one can define quantum walks on interacting Fock spaces using quantum coins and step operators that form the unitary operators of evolution. 


Our contribution in this work is to provide an interacting Fock space perspective of anyons as an IFS generates a quantum field theory from a classical probability measure. This fundamental space induces an algebra that is closed under commutator relation leads to an embedding into a universal enveloping algebra and thus enables us to leverage the Verma modules of the representation in formulating the problem in the language of conformal field theory. Along with the modular invariance of the self dual distance regular graphs, the anyons are characterized by CFT and the closely related mathematical objects to describe duality and fusion, Reidemeister moves, are invariants of knots, braids, and modular data of anyons. We illustrate the comprehensive description of the well known anyons (Ising and Fibbonacci) with symmetry described by K-truncated $SU(2)$ using assocation schemes of Hamming graphs. As quantum groups (q-deformed SU(2), q a root of unity) are for many purposes K-truncated $SU(2)$ and especially in describing CFTs \cite {Sierra1989} we achieve the description of Ising and Fibbonacci anyons by considering the algebra generated by the creation, number, and annihilation operators of the IFS for Hamming graph and using its closure with respect to commutation to induce an enveloping algebra and finally embedding it in $SU_q(2)$. This will enable us to use the representation, in terms of Verma modules, of $SU_q(2)$ in the Bose-Mesner algebra of Hamming graph to describe the fusion rules \cite {Fuchs1994}. Apart from the IFS perspective, we leverage existing results to describe anyons in concrete related mathematical objects as an alternate to abstract categorical formulation. 

Accardi \cite {Accardi2017} and his collaborators developed IFS and quantum probability with classical random variables and orthogonal polynomials characterized by Jacobi coefficients. A different choice of Jacobi sequence leads to IFS for Boson, Fermion, or Free Fock spaces. By selecting them for specific class of graphs (distance-regular, Cayley, and automorphism group that are abelian), we get anyon systems. We illustrate this point by stringing together different algebraic frameworks to provide a comprehensive description of anyons in terms of association schemes. We start with a type II matrix $W$ that respects all the three Reidemeister moves that encode crossings of a spin model. The spin model $W$ encodes the partition function of a statistical system and when the crossings are part of a tangle, the compositions vertically and horizontally are expressed in terms of regular and schur matrix multiplications of associations schemes. The matrix $W$ is a member of the Bose-Mesner algebra and this guarantees expressing them as linear combination of the classes that form the basis of an association scheme. The coefficients of the linear combination form the diagonal matrix $T$ of the modular invariant of a chiral conformal field theory. The common eigenvectors of the association scheme form the $S$ matrix, and classes are non negative integers forming the modular data. Alternately, we can start with a non-bipartite distance-regular graph of a self-dual association scheme and synthesize the $W$ matrix by algebraically satisfying the conditions for Reidemeister moves. The matrix $W$ induces a subfactor leading to a Jones tower of algebras whose projections satisfy the braiding relationships. This is a concrete representation as opposed to abstract categorical formulations and as the schemes support modular data of a conformal field theory, we get the connection to rational conformal field theory. In the following sections we describe the notions with examples and along with our earlier publications \cite {Radbalu2020, Radbalu2021} we complete the association schemes-centric picture of anyons and the computational models (unitaries via braids, quantum walks, and quantum Markov chains). 

\section {von Neumann algebras and subfactors}
In this section we briefly recall well established results in subfactors theory. von Neumann algebras of interest are algebras of bounded linear operators of the form $M \subset \mathscr{B}(\mathscr{H})$, where $\mathscr{H}$ is a separable Hilbert space. Our focus is on factors, that is algebras with trivial centers, satisfying the property $M \cap M' =  \mathbb{C}I$. The representation theory of factors is very rich and exotic, and we narrow our interest to that of subfactors. For example, for factors of type I $M_n(\mathbb{C})$ of $n$- dimensional complex matrices, one can think of a representation in terms of commutants as $L^2 (M, tr) = \mathbb{C}^{n^2} = \mathbb{C}^n \otimes \mathbb{C}^n$ with $$\mathscr{B}(L^2 (M, tr)) = M_{n^2}(\mathbb{C}) = \underbrace {M_n (\mathbb{C})}_M \otimes \underbrace {M_n (\mathbb{C})}_{M'}.$$ This is an M-module which is a Hilbert space $\mathscr{H}$ for which we can define the “coupling constant” or the $M$-dimension of $\mathscr{H}$ as a positive real number $dim_M \mathscr{H} \in [0, \infty)$. The M-dimension of the M-module is a discrete number and for any type I factor endowed with an unique trace it is one but there are factors for which the M-dimension can be any positive real number that induces a rich representation theory. One of the easiest ways to construct type $II_1$ factors is to use a group $\mathcal{K}$ with infinite conjugate classes (i.c.c) as $\mathcal{L}(K)$ acting on the module $\mathcal{L}^2(\mathcal{K})$.

\begin {defn} {Jones tower} The tower of algebras constructed by Jones provide projections with a representations of the Braid group. Let $N \subset M$ be type $II_1$ factors and denote the orthogonal projection onto $L^2(N)$ by $$e_N: L^2(M) \rightarrow L^2(N) \subset L^2(M).$$ The next tower in the construction is a double commutant as $$M_1 = (M \cup \{e_N\}){"} \subset \mathscr{B}(L^2(M)).$$ Here $M_1$ is the von Neumann algebra generated by $M$ and $e_N$ in $\mathscr{B}(L^2(M))$. We denote $M_1 = \langle M, e_N \rangle$ and refer to $M_1$ as the basic construction for $N \subset M$. We get the Jones tower by repeating the process as $N \subset M \subset M_1 \subset M_2 \dots$.
\end{defn}

We consider two factors $M$ and $N$ of type $II_1$ with one embedded in another as $N \subset M$. The Jones index for the subfactor $N$ of $M$ is defined as $[M : N] = dim_N L^2 (M)$. An important result of Jones characterizes the possible values the index can take in the continuum above the value 4 and quantized below it.
\begin {theorem} \cite {Jones1983} Let $N \subset M$ be a $II_1$ subfactor, then the possible values of $[M : N]$ are given by $$\{ 4cos^2 (\pi /n): n = 3,4, \dots \} \cup [4, \infty].$$
\end {theorem}
A subfactor with index $\lambda = [M : N]$ induces a Temperley-Lieb algebra $TL(\lambda)$ with generators $e_i$s satisfying the relation:
\begin {align*}
e_i^2 = e_i = e_i^*. \\
e_i e_j = e_j e_i |i - j| \ge 2. \\
e_i e_{\pm 1} e_i = \lambda^{-1}e_i.
\end {align*}
A Temperley-Lieb algebra $TL(\lambda)$ has a representation in the Braid group connecting subfactors to quantum computation.
\
\section{Association schemes}
\begin{figure}
\includegraphics[width=\columnwidth]{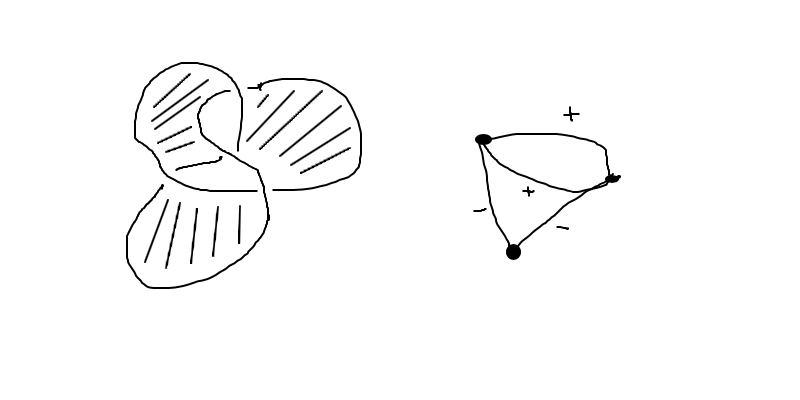}
\caption{ \label{fig:Knots2Graphs} Projection of 3D knot on a plane with adjacent regions colored differently to build a graph with signed edges.}
\end{figure}

V.F.R. Jones constructed planar graphs out of oriented 3D knots by coloring the adjacent regions of their projections with different colors \cite {Jones1989}, and the outer region is colored white. He assigned vertices to dark regions and signed the edges based on the crossings from the left or right (Figure \ref {fig:Knots2Graphs}). He then constructed a statistical ensemble of spins located at the vertices of the graphs with the number of states ranging from two forming the Potts models. The adjacency matrices serving the role of transfer matrices of statistical mechanics are the building blocks of the partition function. The matrices can encode type II and III Reidemeister moves that can be described algebraically, for example if two vertices share edges oppositely signed, they can be removed without affecting the partition function, are related to association schemes that we will look at in detail. Braid relations also arise from these weighted matrices denoted as $(W^+, W^-)$ based on the signs of the edges via subfactor constructions. 

\begin{figure}
\usetikzlibrary {graphs} 
\tikz
  \graph [nodes={draw, circle}, clockwise, radius=.75cm, empty nodes, n=8] {
    subgraph C_n [name=inner] <->[shorten <=1pt, shorten >=1pt]
    subgraph C_n [name=outer]
  };
 \end{figure} 
To describe the topological process in algebraic terms, we can consider an association scheme \cite {Radbalu2020} which is either a collection of adjacency matrices of graphs with a common set of $|\mathfrak{X}| = d$ vertices or it encodes 1-distance, 2-distance, ..., d-distance matrices of the same graph.
Let $X$ be a (finite) vertex set, and let $\mathfrak{X} = \{A_j\}_{j=0}^d$ be a collection of $X\times X$ matrices with entries in $\{0,1\}$. We say that $\mathfrak{X}$ is an \emph{association scheme} if the following hold:
\begin{enumerate}[(1)]
\item $A_0 = I$, the identity matrix;
\item $\sum_{j=0}^d A_j = J$, the all-ones matrix (In other words, the $1$'s in the $A_j$'s partition $X\times X$);
\item For each $j$, $A_j^T \in \mathfrak{X}$; and
\item For each $i,j$, $A_i A_j \in \spn\mathfrak{X}$.
\end{enumerate}
A \emph{commutative} association scheme also satisfies
\begin{enumerate}[(1)]
\setcounter{enumi}{4}
\item For each $i,j$, $A_i A_j = A_j A_i$.
\end{enumerate}
Let us consider an association scheme $\{A_j\}_{j=0}^d$ that is commutative. Then, by the spectral theorem, the matrices $A_0,\dotsc,A_d$ are simultaneously diagonalizable. In other words, the adjacency algebra $\mathscr{A}$ has an alternative basis $E_0,\dotsc,E_d$ of projections onto the maximal common eigenspaces of $A_0,\dotsc,A_d$. Since $\mathscr{A}$ is closed under the Hadamard product forming a Bose-Mesner algebra, there are coefficients $q_{i,j}^k$ such that
\[ E_i \circ E_j = \frac{1}{|X|} \sum_{k=0}^d q_{i,j}^k E_k \qquad (0 \leq i,j \leq d). \]
The coefficients $q_{i,j}^k$ are called the \emph{Krein parameters} of the association scheme. This leads to a commutative hypergroup. Let $m_j = \rank E_j$, and define $e_j = m_j^{-1} E_j$. Then
\[ e_i \circ e_j = \frac{1}{|X|}\sum_{k=0}^d \left( \frac{m_k}{m_i m_j} q_{i,j}^k \right) e_k. \]

The dual notion to Krein parameters, the Intersection numbers $p^k_{ij}$ is defined in terms of matrix product
$A_i \bullet A_j= \sum_{k} p^k_{ij}A_k$.
 The Krein parameters and intersection numbers coincide for the self-dual association schemes that support a spin model $W$.

 \begin {example}
 The Hamming scheme $H(n, q)$ used in coding theory is a useful and relevant example for our discussions on modular invariants. The scheme has vertices n-tuples of elements of a set $Q$ with size $q$. Two tuples $\alpha$ and $\beta$ are adjacent in the graph $X_i$ if they differ in exactly $i$ positions, in other words if they are at Hamming distance $i$.
 \end {example}

\section {Interacting Fock spaces}
Interacting Fock spaces based on non-commutative probability are a specific generalization of symmetric and anti-symmetric Fock spaces relevant in quantum optics and graph theory. In contrast to classical probability theory, there are several different formulations of stochastic independence in the quantum context. This independence is required to define graph products, and based on the monadic operation, different stochastic independence arise that lead to various versions of quantum central limit theorems. In a quantum probability space $(\mathscr{A}, \phi)$ the usual commutative independence $(\phi(bab) = \phi(a)\phi(b^2); a,b \in \mathscr{A})$ such as the one assumed in quantum optics leads to conjugate Brownian motions (measured as quadratures) in the limit. The monotone independence $(\phi(bab) = \phi(a)\phi(b)^2; a,b \in \mathscr{A})$ that is relevant in quantum walks leads to arcsin-Brownian motion (double-horn distribution) aymptotically, and the other two are free and Boolean independences not focused in this work. In the graph context, the independence notions are defined in terms of products of graphs.
\begin {defn} Let $o$ be the fixed vertex of the subconstituent algebra $T$, of complex valued functions defined on the vertices, endowed with an inner product $\langle . , . \rangle$ and a pure state is a linear functional satisfying $$\rho_o (a) = \langle \delta_o , a\delta_o \rangle, a \in T$$.
\end {defn}
\begin {defn} \cite {Obata2007} An IFS associated with the Jacobi sequences $\{\omega_n\}, (\omega_m = 0) \Rightarrow \forall n \ge m, \omega_n = 0, \{\alpha_n\}, \alpha_n \in \mathbb{R}$ is a tuple $(\Gamma(\mathcal{G}) \subset \mathscr{H}, \{\Phi_n\}, B^+, B^-, B^\circ)$ where $\{\Phi_n\}$ are orthogonal polynomials and $B^\pm \Phi_n$ spans $\Gamma$, the subspace of the Hilbert space $\mathscr{H}$. The mutually adjoint operator $B^+, B^-$ and $B^\circ$ satisfy the relations
\begin {align} \label {eqn: FockOperators}
B^+ \Phi_n = \sqrt{\omega_{n+1}} \Phi_{n+1}. \\
B^- \Phi_n = \sqrt{\omega_n} \Phi_{n-1}; B^- \Phi_0 = 0. \\
B^\circ \Phi_n = \phi_n.
\end {align}
\begin {equation} 
xP_n (x) = P_{n+1} (x) + \omega_n P_{n-1} (x) + \alpha_{n+1} P_n (x).
\end {equation}
\end {defn}
With the above IFS, we can associate a graph with an adjacency matrix 
$T = \begin{bmatrix} \alpha_1 & \sqrt{\omega_1}  \\
\sqrt{\omega_1} & \alpha_2 & \sqrt{\omega_2}  \\
& \sqrt{\omega_2} & \alpha_3 & \sqrt{\omega_3}  \\
 & & \ddots & \ddots & \ddots & \\
& & & \sqrt{\omega_{n-1}} & \alpha_n & \sqrt{\omega_n} \\
  & & & & \ddots & \ddots & \ddots 
\end {bmatrix}$
that has the quantum decomposition $T = B^+ + B^- + B^\circ$. The sequence $\{\Phi_n\}$ represents fixing a vertex and stratifying (partitioning based on distance from the fixed vertex) the graph with V set of vertices. Let us fix the Hilbert space $\mathscr{H} = l^2(V)$ of the graph for the rest of the section. The above Jacobi matrix has the structure of a next-neighbor hopping Markov chain and this fact will be used later to make the connection to subfactors.

Let us consider an IFS for the association scheme with d classes $\mathfrak{X}:=\{B_j\}_{j=0}^d$ similar to the development of multi-dimensional orthogonal polynomials by Accardi \cite{Accardi2017}. For example, association schemes induced by finite cyclic groups are commutative, and as the schemes are self-duals and as a consequence support spin models \cite {Nomura1998-1}. In our case the variables are matrices, and the algebra is also closed under Schur multiplication $\circ$ and thus a *-algebra with the ladder operators (CAPs) of the IFS can be defined in terms of the parameters of the association scheme. These are positive real numbers and can be normalized to become a probability measure with their square root interpreted as probability amplitudes. We will work with the induced hypergroup of the conjugacy classes where the Krein numbers can be interpreted as collision probabilities. The classes of the association schemes are referred to as modes, and they represent different graphs with common vertices. Orthogonal polynomials in finite number of variables were treated in \cite {Stan2004} and the commutation relations between the ladder operators derived. Now, we can see that an IFS is a T-algebra endowed with a quantum probability space in a pure state. This connection is important as it lets us lift the result that the association schemes with spin models have their T-modules thin \cite {Curtin1999} to the IFS framework. This result is a necessary condition for an association scheme to support a spin model which is to have every T-module $T(x), \forall x \in X$ to be thin, and of dimension less than or equal to one, which is a very stringent condition. This implies that we have to look for an IFS whose disjoint union of Hilbert spaces are thin with respect to all the pure states by adjusting the Jacobi parameters. This opens up ways to investigate systems with modular invariance using the tools of classical probability and orthogonal polynomials.

\begin {example} Let us now define the IFS for a binary Hamming graph H(d, 2) \cite {Obata2007} of d-dimension that we will use later to establish our main result. The IFS for a distance-regular graph is the tuple $(\Gamma(\mathcal{G}) \subset \mathscr{H}, \{\Phi_n\}, B^+, B^-, B^\circ)$, where $(\mathcal{G})$ is a distance-regular graph and $\Phi_n$ are the stratifications of the graph. The annihilation, creation, and number operators satisfy the relations \eqref {eqn: FockOperators}. For the distance-regular graphs the relations are cast in terms of intersection numebrs as:
\begin {align} 
B^+ \Phi_n = \sqrt{p_{1,n}^{n+1}p_{1,n+1}^{n}} \Phi_{n+1}. \\
B^- \Phi_n = \sqrt{p_{1,n-1}^{n}p_{1,n}^{n-1}} \Phi_{n-1}; B^- \Phi_0 = 0. \\
B^\circ \Phi_n = p_{1,n}^n\phi_n.
\end {align}
\begin {equation} 
xP_n (x) = P_{n+1} (x) + \omega_n P_{n-1} (x) + \alpha_{n+1} P_n (x).
\end {equation}
In the case of the binary Hamming graph in d-dimensions the intersection numbers are:
\begin {align*}
p_{1,n}^n &= n (d - 2). \\
p_{1,n}^{n-1} &= (d - n + 1). \\
p_{1, n-1}^n &= n.
\end {align*}
\end {example}
We will use this setup to establish our result on embedding the Bose-Mesner algebra into a quantum group.
\section {Type-II matrices}

\begin {defn}
If $M \circ N = J$, we say that $N$ is the Schur inverse of $M$, and denote it $M^-$. A type-II matrix \cite {Nomura1998} is a Schur invertible $n \times n$ matrix $W$ over $\mathbb{C}$ such that $WW^{(-)^T} = nI$ that can be thought of as a specific generalization of Hadamard matrices where $n$ denotes the number of spin states. In other words these matrices satisfy Reidemeister moves of Types II. Other examples of type-II matrices include character tables of abelian groups. For a type-II matrix $W$, the matrix $\frac{1}{\sqrt{n}}W$ is unitary with the absolute value of each entry is one which is of importance to us in this work. In addition if the matrix satisfies the following condition then it encodes type III Reidemeister move that are required for braid relations:
\begin {equation} \label {TypeIIIEqn}
\sum_{y \in X} \frac {W_{ay} W_{by}} {W_{xy}} = \sqrt{|X|} \frac {W_{ab}} {W_{xa} W_{xb}}.
\end {equation}
\end {defn}

\begin {defn} Let $W$ be a Schur-invertible $n \times n$ matrix. We define $\mathscr{N}_W$ to be the set of matrices for which all the vectors $W e_i \circ W^{(-)} e_j, 1 \leq i,j \leq n$ (that is Schur multiplication of the corresponding columns of $W$ and its Schur-inverse) are all eigen vectors. The  matrix $W_4$ is of type I, and the rest are of type II corresponding to Reidemeister moves respectively.
\end {defn}

\begin {example}
We start with an example of a type II matrix satisfying equation \eqref{TypeIIIEqn} induced by an association scheme with two classes $\{I = \begin{bmatrix} 1 & 0 & 0 \\ 0 & 1 & 0 \\ 0 & 0 & 1\end {bmatrix}, \text {  and  } B = (J - I), \text { where } J = \begin{bmatrix} 1 & 1 & 1\\ 1 & 1 & 1 \\ 1 & 1 & 1 \end {bmatrix}$.
$$W = c(I - tB), c^2 = \frac {\sqrt{n}} {1 - (n - 1)t}, t + t^{-1} = n - 2 $$.
\end {example}
This example illustrates how we can synthesize a type II matrix encoding the partition function of a spin system from an association scheme called the 3-state Potts model. Solving these equations may not always be possible except for the associations schemes that support spin models.

\begin {example} {Type II matrices and algebraic equations for Reidemeister moves}
$$
W_1 = \begin{bmatrix} 1 & 1 \\ 1 & -1 \end {bmatrix}; W_1^{(-)} = \begin{bmatrix} 1 & 1 \\ 1 & -1 \end {bmatrix}.$$
$$W_2 = \begin{bmatrix} 1 & \omega & 1 \\ \omega & 1 & 1 \\ 1 & 1 & \omega \end {bmatrix}; W_2^{(-)} = \begin{bmatrix} 1 & \omega^2 & 1\\ \omega^2 & 1 & 1 \\ 1 & 1 & \omega^2\end {bmatrix};  \omega \text{ is cube root of unity}.$$
$$
W_3 = \begin{bmatrix} 1 & 1 & 1 & 1\\ 1 & 1 & -1 & -1 \\ 1 & -1 & \omega & -\omega \\ 1 & -1 & -\omega & \omega \end {bmatrix}; W_3^{(-)} = \begin{bmatrix} 1 & 1 & 1 & 1\\ 1 & 1 & -1 & -1 \\ 1 & -1 & \omega^2 & -\omega^2 \\ 1 & -1 & -\omega^2 & \omega^2 \end {bmatrix};$$

\end {example}


We can get braiding relations from subfactor construction as we detail in the next section as that will provide a modular invariance \cite {Gann2005}. This is because type II matrices have to satisfy an additional condition, invariance with respect to a Reidemeister move III, to support a spin model and thus a modular invariance. 

\begin {example}
Let $$Y_{ab} (x) = \frac {W_2(x, a)}{W_2(x, b)}$$ and by fixing b we can get the eigenspace of the scheme as $$\left(1, 1, 1 \right)^T;\left( 1/\omega, \omega, 1 \right)^T;\left(1, \omega, 1/\omega \right)^T $$.

It is easy to verify that the above three vectors form the eigen space for the association scheme generated by the group $\{1, \omega, \omega^2  \}$  with the class 
\begin {equation*}
A_1 = \begin{bmatrix} 1 & 0 & 0 \\ 0 & 1 & 0 \\ 0 & 0 & 1 \end {bmatrix}; A_2 = \begin{bmatrix} 0 & 0 & 1\\ 1 & 0 & 0 \\ 0 & 1 & 0\end {bmatrix}; A_3 = \begin{bmatrix} 0 & 1 & 0\\ 0 & 0 & 1 \\ 1 & 0 & 0\end {bmatrix}.
\end {equation*}

\begin{figure}[htbp]
 \centering
\begin{tikzpicture}[node distance =0.85 cm and 0.85 cm]
\draw[step=1cm,gray,very thin] (-1.9,-0.9) grid (1.9,2.9);
\node[draw,circle,inner sep=4pt,fill=red] at (-1, 0) {};
\node[draw,circle,inner sep=4pt,fill=blue] at (-1, 1) {};
\node[draw,circle,inner sep=4pt,fill=green] at (-1, 2) {};
\node[draw,circle,inner sep=4pt,fill=blue] at (0, 0) {};
\node[draw,circle,inner sep=4pt,fill=green] at (0, 1) {};
\node[draw,circle,inner sep=4pt,fill=red] at (0, 2) {};
\node[draw,circle,inner sep=4pt,fill=green] at (1, 0) {};
\node[draw,circle,inner sep=4pt,fill=red] at (1, 1) {};
\node[draw,circle,inner sep=4pt,fill=blue] at (1, 2) {};


\end{tikzpicture}
\caption{Distance-regular graph corresponding to the above association scheme with matrices $A_1, A_2, A_3$ in colors green, blue, red, black respectively.}
\label{fig:AS-2}
\end{figure}
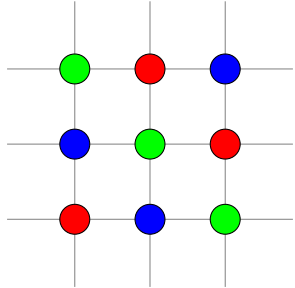 
With $W_2$ as the type-II matrix, we can generate braiding relations through induced subfactors and thus a system of anyons.

For an example of an association scheme with four classes let $$Y_{ab} (x) = \frac {W_3(x, a)}{W_3(x, b)}$$ and by fixing b we can get the eigenspace of the scheme as $$\left(1, 1, 1, 1 \right)^T;\left( 1, 1, -1, -1 \right)^T;\left(1, -1, i, -i \right)^T;\left(1, -1, -i, i \right)^T $$ (Figure \ref {fig:AS-3}).

It is easy to verify that the above three vectors form the eigenspace for the association scheme generated by the multiplicative group $\{1, -1, i, -i  \}$  (if matrix $W_3$ has other complex numbers a group can be formed in a similar fashion using the corresponding inverse) with the class of adjacency matrices and the corresponding distance-regular graph \cite {vidali2018}:
\begin {equation*}
A_1 = \begin{bmatrix} 1 & 0 & 0 & 0\\ 0 & 1 & 0 & 0\\ 0 & 0 & 1 & 0 \\ 0 & 0 & 0 & 1 \end {bmatrix}; A_2 = \begin{bmatrix} 0 & 1 & 0 & 0\\ 1 & 0 & 0 & 0\\ 0 & 0 & 0 & 1 \\ 0 & 0 & 1 & 0\end {bmatrix}; A_3 = \begin{bmatrix} 0 & 0 & 1 & 0\\ 0 & 0 & 0 & 1\\ 1 & 0 & 0 & 0 \\ 0 & 1 & 0 & 0\end {bmatrix}; A_4 = \begin{bmatrix} 0 & 0 & 0 & 1\\ 0 & 0 & 1 & 0\\ 0 & 1 & 0 & 0 \\ 1 & 0 & 0 & 0\end {bmatrix}.
\end {equation*}
Chan and Godsil \cite {Chan2010} provide more examples for the construction of type II matrices in a systematic manner. As there are stringent conditions to satisfy type III moves, the number anyon systems will be limited. 
With $W$ we can also build a subfactor to have a representation of braids expanding the family of anyon systems that can be considered.

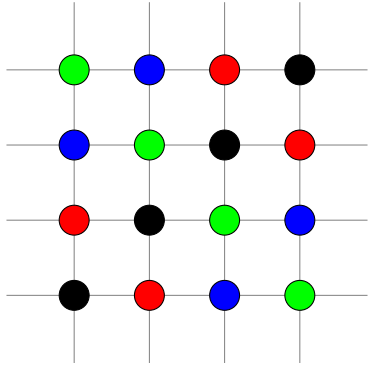
\begin{figure}[htbp]
 \centering
\begin{tikzpicture}[node distance =0.85 cm and 0.85 cm]
\draw[step=1cm,gray,very thin] (-1.9,-1.9) grid (2.9,2.9);
\node[draw,circle,inner sep=4pt,fill=black] at (-1, -1) {};
\node[draw,circle,inner sep=4pt,fill=red] at (-1, 0) {};
\node[draw,circle,inner sep=4pt,fill=blue] at (-1, 1) {};
\node[draw,circle,inner sep=4pt,fill=green] at (-1, 2) {};
\node[draw,circle,inner sep=4pt,fill=red] at (0, -1) {};
\node[draw,circle,inner sep=4pt,fill=black] at (0, 0) {};
\node[draw,circle,inner sep=4pt,fill=green] at (0, 1) {};
\node[draw,circle,inner sep=4pt,fill=blue] at (0, 2) {};
\node[draw,circle,inner sep=4pt,fill=blue] at (1, -1) {};
\node[draw,circle,inner sep=4pt,fill=green] at (1, 0) {};
\node[draw,circle,inner sep=4pt,fill=black] at (1, 1) {};
\node[draw,circle,inner sep=4pt,fill=red] at (1, 2) {};
\node[draw,circle,inner sep=4pt,fill=green] at (2, -1) {};
\node[draw,circle,inner sep=4pt,fill=blue] at (2, 0) {};
\node[draw,circle,inner sep=4pt,fill=red] at (2, 1) {};
\node[draw,circle,inner sep=4pt,fill=black] at (2, 2) {};

\end{tikzpicture}
\caption{Graph corresponding to the above association scheme with matrices $A_1, A_2, A_3, A_4$ in colors green, blue, red, black respectively.}
\label{fig:AS-3}
\end{figure} 
\end {example}
\begin {example} \cite {vidali2018}
Let us consider the 2D graph with vertices colored describing a generalized Hamming scheme shown in figure \ref{fig:AS-1} whose adjacency matrices can be shown to form a Bose-Mesner algebra.
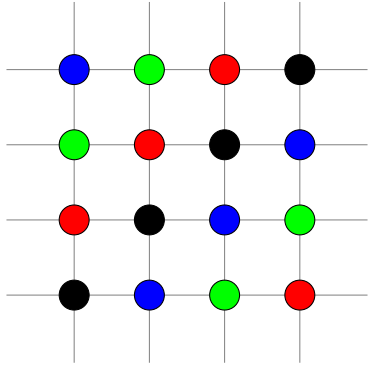
\begin{figure}[htbp]
 \centering
\begin{tikzpicture}[node distance =0.85 cm and 0.85 cm]
\draw[step=1cm,gray,very thin] (-1.9,-1.9) grid (2.9,2.9);
\node[draw,circle,inner sep=4pt,fill=black] at (-1, -1) {};
\node[draw,circle,inner sep=4pt,fill=red] at (-1, 0) {};
\node[draw,circle,inner sep=4pt,fill=green] at (-1, 1) {};
\node[draw,circle,inner sep=4pt,fill=blue] at (-1, 2) {};
\node[draw,circle,inner sep=4pt,fill=blue] at (0, -1) {};
\node[draw,circle,inner sep=4pt,fill=black] at (0, 0) {};
\node[draw,circle,inner sep=4pt,fill=red] at (0, 1) {};
\node[draw,circle,inner sep=4pt,fill=green] at (0, 2) {};
\node[draw,circle,inner sep=4pt,fill=green] at (1, -1) {};
\node[draw,circle,inner sep=4pt,fill=blue] at (1, 0) {};
\node[draw,circle,inner sep=4pt,fill=black] at (1, 1) {};
\node[draw,circle,inner sep=4pt,fill=red] at (1, 2) {};
\node[draw,circle,inner sep=4pt,fill=red] at (2, -1) {};
\node[draw,circle,inner sep=4pt,fill=green] at (2, 0) {};
\node[draw,circle,inner sep=4pt,fill=blue] at (2, 1) {};
\node[draw,circle,inner sep=4pt,fill=black] at (2, 2) {};
\end{tikzpicture}
\caption{Graph corresponding to the above association scheme with matrices $A_1, A_2, A_3, A_4$ in colors blue, green, red, black respectively.}
\label{fig:AS-1}
\end{figure} 
In this graph, the 16 vertices form the connectivity of the first class of the scheme which is represented by the identity matrix $A_0 = \mathbb{I}_{16\times 16}$. The connectivity between two vertices $(u, v)$ is defined if they are both in the same column or row of the lattice as the lines here denote cliques. As we can see each vertex is connected to three other vertices in the same row and three others within the same column as the graph consists of s a four interconnected tetrahedrons with connectivity represented on a plane. So, the matrix $A_1$ will have exactly six non zero entries in each row and column. The connectivity between two vertices for the matrix $A_3$ is defined if they both are of the same color. So matrix $A_2$ will have three non zero entries in each column and row. If two vertices are not connected in the previous sense then they are connected are form adjacency for $A_4$ and this matrix will have exactly six non zero entries in each column and row. Now, we can understand the intersection numbers starting with $p_{00}^0$ that comes from the product of two identity matrices on the left and an identity matrix on the right, and so $p_{00}^0 = 1$. Similarly, we can see the other entries of $(p^0_{ij})_{i,j=0}$ follows from the entries of $A_i s$. It is easy to work out the entries of the other three matrices in a similar fashion. 
\begin {align*}
(p_{ij}^0)_{i,j=0}^3 &= \begin{bmatrix} 1 & 0 & 0 & 0\\ 0 & 6 & 0 & 0\\ 0 & 0 & 3 & 0 \\ 0 & 0 & 0 & 6 \end {bmatrix}; (p_{ij}^1)_{i,j=0}^3 &= \begin{bmatrix} 0 & 0 & 1 & 0\\ 0 & 2 & 0 & 4\\ 1 & 0 & 2 & 0 \\ 0 & 4 & 0 & 2\end {bmatrix}; \\ (p_{ij}^2)_{i,j=0}^3 &= \begin{bmatrix} 0 & 1 & 0 & 0\\ 1 & 2 & 1 & 2\\ 0 & 1 & 0 & 2 \\ 0 & 2 & 2 & 2\end {bmatrix}; (p_{ij}^3)_{i,j=0}^3 &= \begin{bmatrix} 0 & 0 & 0 & 1\\ 0 & 2 & 2 & 2\\ 0 & 2 & 0 & 1 \\ 1 & 2 & 1 & 2\end {bmatrix}.
\end {align*}
This scheme is self-dual as the intersection numbers and Krein parameters coincide as computed by the sage package \cite {vidali2018}. That means there is a linear map $\psi$ from the Bose-Mesner algebra $\mathcal{A}$ to itself such that \cite {Jaeger1996}
\begin {align*}
\psi(\psi(M)) &= |X| M, M \in \mathcal{A}. \\
\psi (MN) &= \psi(M) \circ \psi (N), \text {  relating regular and Schur matrix products.}
\end {align*}
Now we can define $W^- = \sum_i t_i A_i, t_i \neq t_j$ and 

now, we can define the parameters $(a, q)$ of the spin model from
\begin {align*}
\mathbb{I} \circ W^- &= a^{-1}\mathbb{I}. \\
\mathbb{I} \circ W^+ &= a \mathbb{I}. \\
\mathbb{J}W^+ &= W^+\mathbb{J} = qa^{-1}\mathbb{J}.
\end {align*}
This spin models has the modular invariance property as $$(PT)^3 = q^3 a^{-1}\mathbb{I}.$$ where the matrix $P$ has the eigenvectors of the scheme as columns, and the diagonal matrix $T$ has $t_i$ as elements.
\end {example}
\
\begin {example}
Generalized Hamming schemes can be constructed from any association scheme $\mathscr{A}$ as $H(n, \mathscr{A})$ \cite {Godsil2010}. Let $\mathscr{A}$ be an association scheme with $d$ classes and vertex set $V$. If $v,w$ are elements of $V^n$, let $h(v,w)$ be the vector of length $d + 1$ with $rth-entry$ equal to the
number of coordinates $j$ such that $v_j$ and $w_j$ are $r$-related in $\mathscr{A}$. For any $n$-
tuples $v$ and $w$, the vector $h(v,w)$ has non-negative integer entries, and these
entries sum to $n$. Conversely, any such vector can be written as $h(v,w)$ for
some $v$ and $w$. If $x$ is an integer vector of length $d+1$ with entries summing
to $n$, let $A_x$ be the $01$-matrix with rows and columns indexed by $V_n$, and
with $\mathscr{A}_{v,w}$ equal to one if and only if $h(v,w) = x$. This set of matrices $H(n,\mathscr{A})$ forms an association scheme.
\end {example}
\begin {lemma} \cite {Godsil2010} If the association scheme $\mathscr{A}$ satisfies the modular invariance property, so does $H(n,\mathscr{A})$.
\end {lemma}
Using the above lemma, we can construct several examples of Hamming association schemes with the modular invariance property. 
\section {Commuting Squares}
Let $D_X$ consist of all diagonal matrices in $M_X$, and let $W$ be an invertible type II matrix in $M_X$ where $X$ is the set of vertices of the graph representing the association scheme. Then, the following is a commuting square from which a hyperfinite $II_1$ subfactor may be obtained from the above commuting square of finite dimensional C*-algebras via iteration of the basic construction \cite {Nicoara2010}, \cite {Jones1989}.

\begin {align*}
D_X  & & \bigsubset[1.4]   & & M_X  \\
\bigcup  & &  & & \bigcup   \\
\mathbb{C} & & \bigsubset[1.4] & & W^{-1} D_X W
\end {align*}
We can construct from the commuting square the Jones tower of algebras on the top and bottom rows to get two hyperfinite factors and a subfactor as: 
$$A_\infty \simeq C_\infty \simeq \mathcal{R}; C_\infty \subset A_\infty$$.
\begin {align*}
A_0  & & \rightarrow   &A_1 & \dashrightarrow A_\infty  \\
\uparrow  & &  \uparrow & &  \uparrow \\
\mathbb{C}_0 & & \rightarrow &\mathbb{C}_1  & \dashrightarrow \mathbb{C}_\infty
\end {align*}
The basic construction of the tower of algebras can proceed vertically as well. The limit algebras and hence the subfactor can be constructed with finite dimensional ones (matrices) by applying Ocneanu{'}s compactness argument (\cite {Ocneau1991}, Theorem 11.15 \cite {Evans1998}) giving us the Temperley-Lieb planar algebra $$\bigcup_k P_K, P_K = A{'}_{\infty 0}\bigcap A_{\infty k} = A{'}_{1 0}\bigcap A_{0 k}$$ \cite {Evans2023}.  The 2D lattice of algebras is the source of extracting subfactors with Jones index $4cos^2(\pi / (k + 1))$.  

Starting from a Type II matrix $W$ induced by a self-dual association scheme, we can construct commuting squares and then a subfactor matching the ones from the $SU(2)_K$ approach. The fusion rules encoded by the matrix $W$ continue in the tower of algebras as at each step we only add the braiding projections. This then leads to a 3D topological quantum field theory using the Reshetikhin-Turaev method \cite {Turaev1991}. Along with this construction when we build a 2D RCFT the bulk-boundary correspondence is established.

Before we map the universal enveloping algebra $sl(2, C)$ algebra onto quantum $U_q sl(2, C)$ we need few definitions and results.
\begin {defn} \cite {Terwilliger2003} Let $V$ denote a vector space over the field $K$ with finite positive dimension. A Leonard pair on $V$ is an ordered pair of linear transformations $A : V \rightarrow V$ and $B : V \rightarrow V$ that satisfy conditions (i), (ii) below:
(i) There exists a basis for $V$ with respect to which the matrix representing $A$ is irreducible tri-diagonal and the matrix representing $B$ is diagonal.
(ii) There exists a basis for $V$ with respect to which the matrix representing $A$ is diagonal, and the matrix representing $B$ is irreducible tri-diagonal.
\end {defn}
\begin {example} A simple example of a Leonard pair is the following two matrices:
$$A = \begin{bmatrix} 0 & 3 & 0 & 0 \\ 1 & 0 & 2 & 0 \\ 0 & 2 & 0 & 1 \\ 0 & 0 & 3 & 0 \end {bmatrix}
B = \begin{bmatrix}
 3 & 0 & 0 & 0 \\ 0 & 1 & 0 & 0 \\ 0 & 0 & -1 & 0 \\ 0 & 0 & 0 & -3
\end {bmatrix}$$
\end {example}

Leonard pairs can be constructed with one of the pair as the diagonal matrix $(d, d -2, \dots, -d)$ and the other using special orthogonal polynomials such as q-Racah and q-Kwatchouk series \cite {Terwilliger2003}. These two matrices form the bases for conformal blocks as we establish next \cite {Reshetikhin1989}. Kaul \cite {Kaul1994} provided the details of the two bases for conformal blocks for 4-spin correlators related by q-Racah coefficients. He also showed how to extend the approach to 8-spins placed at the crossings of a knot. With the Leonard pair construction from representations of $U_q(sl(2, C))$ the number of spins can be arbitrarily large.

 \begin {defn} The Kwatchouk algebra $\mathcal{K}_\omega$ is an algebra over $\mathbb {C}$ generated by A and B satisfying the relations: 
\begin {align*}
A^2 B - 2ABA + BA^2 &= B + \omega A. \\
B^2 A - 2BAB + AB^2 &= A + \omega B.
\end {align*} 
\end {defn}
\begin {lemma} \cite {Huang2023} The algebra $\mathcal {K}_\omega$ has a presentation:
\begin {align*}
C & = [A, B]. \\
[A, C] &= B + \omega A. \\
[C, B] &= A + \omega B.
\end {align*}
\end {lemma}
\begin {theorem} \cite {Huang2023} There is an isomorphism between $U(sl(2, C))$ and $\mathcal{K}_\omega$ when $\omega^2 \neq 1$ as follows:
\begin {align*}
A &\rightarrow (\frac {1 + \omega}{2})B^+ + (\frac {1 - \omega}{2})B^- - (\frac {\omega}{2})H.\\
B &\rightarrow (\frac {\omega}{2})H.\\
C &\rightarrow (\frac {1 + \omega}{2})B^+ + (\frac {1 - \omega}{2})B^-.
\end {align*}
\end {theorem}
We now set up and establish the main result.

\begin {theorem} The IFS of a binary Hamming graph of d-dimension induces a *-algebra, on the Hilbert space $\mathscr{H} = \mathbb{C}^2\otimes L^2(\mathcal{G})$, that is homomorphic to the universal enveloping algebra $U_q sl(2, C)$ where $q$ is a root of unity.
\end {theorem}
\begin {proof}
Let us define the operator $$H = [B^+, B^-] = B^+ B^- - B^- B^+.$$ For the case of binary Hamming graph this computes to 
$$H \Phi_n = (\omega_n - \omega_{n+1})\Phi_n = \frac {(2n - d)}{d} \Phi_n.$$
Now, we have the commutator $$[H, B^+] = \frac {2(n + 1) -d - 2n + d}{d}\sqrt{\omega_{n+1}} = 2B^+$$. Similarly, we compute $$[H, B^-] = -2B^-.$$ The algebra generated by $(H, B^+, B^-)$ satisfy the same commutator relationship as $sl(2, C)$ and so we can embed into the universal enveloping algebra $U(sl(2, C))$ which is a Hopf algebra.



The isomorphism between $K_\omega$ algebra and $U(sl(2, C))$ leads to Leonard pairs A and B that are the generators of $\mathcal{K}_\omega$ algebra acting on the modules of $U(sl(2, C))$. We can define finite irreducible modules of $U_q sl(2, C)$ \cite {Kessel1995} to which we can associate Leonard pairs of 
quantum q-Krawtchauk polynomials. This association between $U_q sl(2, C)$ modules and Leonard pairs is stated in \cite {Terwilliger2003} for values of q that are not roots of unity as that would make two eigen values identical. However, the result still holds when q is a root of unity if we carefully choose $d$ as we explain later.

Therefore, we can map modules of Hamming graphs into $U_q sl(2, C)$ \cite {Fairlie1990} enabling us to use the Verma modules that are representations of the larger algebra into our IFS induced ones. 

The K-truncated $SU(2)$ algebras are related to the quantum groups $U_q(2, C)$ by the relation $q =exp {\frac {2\pi i}{k + 2}}$. For describing different anyons, for Ising k = 2 and Fibonacci k = 3, the Hamming, with the strata large enough to hold the association between Leonard pairs and the quantum group modules, algebra can embedded into the quantum algebra by choosing the value of q as a root of unity and satisfying the above expression. There is an additional constraint in setting up the Leonard pairs compatible with the $U_q(2, C)$  modules as the eigen values of the pair all have to be distinct. That is (Example 6.3 \cite {Terwilliger2003}) $$\theta_i =\frac {\epsilon q^{d - 2i}} {q - q^{-1}}, 0\leq i \leq d, \epsilon \in \{-1, 1\}$$ with the constraint that the root exponent cannot be $1 \leq m \leq d$ as that would make the eigenvalue same as the case where $d = 2i$. We start with $k$ to determine $q$ and based on that determine $d$ which is the number of strata in Hamming graph as odd as that would guarantee distinct eigenvalues.

The above construction of $U_qsl(2, C)$ modules in (Example 6.3 \cite {Terwilliger2003}) can be carried out for any arbitrary $\epsilon \neq 0$ making it as a quotient of Verma module.
\end {proof}
\section {Summary and Conclusions}
We discussed an IFS based perspective for topological quantum systems and used the connection between fusion algebras central to conformal field theory and association schemes to describe anyon systems. We identified the modular data behind schemes and described duality, Reidemeister moves, and braiding with the same concrete mathematical structures in terms of matrices. Finally, we embedded the algebra of Hamming graphs into $U_q sl(2, C)$ \cite {Fairlie1990} enabling us to use the Verma modules that are representations of the larger algebra into our IFS induced ones. This connection to CFT can be used to explore topological phases that would support new paradigms of computation with colored braids and multicolored links in a systematic fashion using algebraic tools.  
\
\section {Acknowledgement} The author is grateful to Paul Terwilliger for suggesting, in a private communication, the conditions under which the Leonard pairs of q-Krawtchauk polynomials have $u_q(2, C)$ modules.

\section {Declarations}
\textbf {Funding and/or Conflicts of interests/Competing interests:} The  funding information is not applicable and there are no conflicting or competing interests associated with this manuscript.\\
\textbf {Data availability:} This manuscript has no associated data.

\bibliographystyle{abbrv}

\end{document}